\documentclass[5p,a4paper]{elsarticle}
\makeatletter
\def\ps@pprintTitle{%
 \let\@oddhead\@empty
 \let\@evenhead\@empty
 \def\@oddfoot{\centerline{\thepage}}%
 \let\@evenfoot\@oddfoot}
\makeatother
\usepackage{lineno}
\usepackage{color}
\usepackage{upgreek}
\usepackage{amssymb}
\usepackage{amsmath}
\usepackage{soul}
\biboptions{numbers,sort&compress}

\renewcommand{\vec}[1]{\mathbf{#1}}

\newcommand{\figref}[1]{Figure~\ref{#1}}

\newcommand{\tabref}[1]{Table~\ref{#1}}
\newcommand{\vecgrk}[1]{\boldsymbol{#1}}
\newcommand{\degreeC}{$^{\circ}\!$C}
\newcommand{\pme}[2]{$#1\!\pm\!#2$}
\definecolor{orange}{rgb}{1,0.5,0}

\begin{document}
\begin{frontmatter}
\title{Probabilistic design of a molybdenum-base alloy using a neural network}
\author{B.D.~Conduit}
\address{Rolls-Royce plc, PO Box 31, Derby, DE24 8BJ, United Kingdom}
\author{N.G.~Jones}
\address{Department of Materials Science \& Metallurgy, University of Cambridge, 27 Charles Babbage Road, Cambridge, CB3 0FS, United Kingdom}
\author{H.J.~Stone}
\address{Department of Materials Science \& Metallurgy, University of Cambridge, 27 Charles Babbage Road, Cambridge, CB3 0FS, United Kingdom}
\author{G.J.~Conduit}
\address{Cavendish Laboratory, University of Cambridge, J.J. Thomson Avenue, Cambridge, CB3 0HE, United Kingdom}
\date{\today}
\begin{abstract}
  An artificial intelligence tool is exploited to discover and characterize
  a new molybdenum-base alloy that is the most likely to simultaneously
  satisfy targets of cost, phase stability, precipitate content, yield
  stress, and hardness.  Experimental testing demonstrates that the proposed
  alloy fulfills the computational predictions, and furthermore the physical
  properties exceed those of other commercially available Mo-base alloys for
  forging-die applications.
\end{abstract}
\begin{keyword}
modeling, refractory metals, forging, mechanical properties, neural network
\end{keyword}
\end{frontmatter}

The contemporary approach to develop new materials is experiment driven
trial and improvement~\cite{Curtarolo13}. This approach may take up to
twenty years to design and verify a new material. The long lead time rules
out designing new materials alongside products, instead forcing engineers to
compromise products around the shortcomings of pre-existing materials. The
opportunity to discover materials computationally has the potential to
empower engineers to design optimal materials at the same time as new
products~\cite{Kuehmann09}, bringing materials into the heart of the design
process. Previous approaches to design new materials on a computer include
ranking compositions with a Pareto
set~\cite{Bligaard03,Greeley06,Lejaeghere13}, characterizing materials with
a principal component analysis~\cite{Toda13}, robust
design~\cite{Backman06}, and the orthogonal optimization of different
properties~\cite{Joo09,Xu09,Reed09,Kuehmann09,Tancret13}. However, these
methods cannot simultaneously optimize the compromise between material
properties and capture the deep correlations between composition and final
properties. Therefore, in this paper, a new artificial intelligence
tool~\cite{Conduit17} that can capture the full composition-property
relationship is used to propose the new Mo-base alloy for forging die
applications that is most likely to satisfy all target properties
simultaneously.

Mo-base alloys offer exceptional strength at high temperature. This makes
them suitable for refractory applications including fission and fusion
reactors, rocket engine nozzles, furnace structural components, and forging
dies.  However, the next generation of forging applications will demand yet
higher operating temperature requiring a new generation of Mo-base
alloys. Existing Mo-base alloys such as MHC (1.1wt\%~Hf, 0.1wt\%~C,
balance~Mo), TZC (1.2wt\%~Ti, 0.1wt\%~C, 0.3wt\%~Zr, balance~Mo), TZM
(0.5wt\%~Ti, 0.02wt\%~C, 0.08wt\%~Zr, balance~Mo), and ZHM (1.2wt\%~Hf,
0.1wt\%~C, 0.4wt\%~Zr, balance~Mo)~\cite{Chang1966} contain minimal
strengthening precipitates, so there is an opportunity to optimize the
content of HfC and other carbides in Mo-base alloys to improve strength at
high-temperature. Critically, the effective exploitation of strengthening
precipitates requires a firm understanding of the relationship that exists
between the alloy composition and it phase stability, strength and cost; a
multidimensional problem that is an ideal application of an artificial
intelligence tool.

The first section of this paper outlines the artificial intelligence tool
and specifies the chosen targets for the relevant material properties: cost,
phase stability, HfC content, yield stress, and hardness.  In the second
section, the tool is used to propose the new Mo-base alloy that is most
likely to exceed the design targets. The final section presents experimental
results for the phase stability, HfC content, and hardness to verify the
model predictions and demonstrate that the alloy has properties that surpass
those of other commercially available Mo-base forging die alloys.

The goal of the neural network tool is to predict the composition and
processing variables that are most likely to produce a material that
fulfills the multi-criteria target specification. The tool and methodology
follows the prescription developed in Ref.~\cite{Conduit17}. The tool first
constructs a predictive model for each property as a function of the
composition, which for the Mo-base alloys presented in this paper comprises
of the elements $\{\rm{Nb},\rm{Ti},\rm{C},\rm{Zr},\rm{Hf},\rm{W},\rm{Mo}\}$.
The tool can then calculate the likelihood that a putative composition
fulfills the target specification, so that it can search composition space
for the alloy most likely to meet the target specification.

\begin{table}
 \small\centering
 \begin{tabular}{lccc}
   Property&Target&Approach&Data\\
   \hline\hline
   Cost&$<$52/cycle&
   Physical&\cite{ati2013,Shields13,ntl2013,rsu2013,AmericanElements2013}\\
   \hline
   Phase stability&$>$81wt\%&
   {\sc calphad}&\cite{mohfcRef,mowcRef,monbcRef,moticRef,mozrcRef,thermocalcgeneral02,SSOL_2016}\\
   HfC\,content&$>$1wt\%&
   {\sc calphad}&\cite{mohfcRef,mowcRef,monbcRef,moticRef,mozrcRef,thermocalcgeneral02,SSOL_2016}\\
   $1000${\degreeC}\,yield\,stress&$>$398MPa&
   Neural\,net&212\cite{barr1959,climax1955,climax1960,imgram1960,mcardle1963,
   neff1961,redden1959,semchyshen1959,
   semchyshen1959_1,semchyshen1959_2,semchyshen1961}\\
   $1000${\degreeC}\,hardness&$>$1908MPa
   &Neural\,net&
   740\cite{barr1959,climax1955,climax1960,imgram1960,mcardle1963,neff1961,
   redden1959,semchyshen1959,semchyshen1959_1,semchyshen1959_2,
   semchyshen1961,bruckart1950,feild1961,foldes1965,foyle1961,glasier1959,
   hall1959,harmon1960,houck1960,Lement1960,marquardt1959,schmidt1963,
   semchyshen1957,sikora1959,sikora1962,wilkinson1969}
 \end{tabular}
 \caption{The approach used to predict properties, the property
   targets, number of experimental points used to train neural network
   models, and references for the data are shown.}
 \label{tbl:DataSources}
\end{table}

Materials must fulfill a wide ranging specification to ensure that they best
meet the needs of their target application. The properties that were
optimized in the design of the Mo-base alloys are shown in
\tabref{tbl:DataSources}. With properties depending on contrasting physics,
for each property a different source of data must be adopted, which are
referenced in the tables. The cost per cycle -- the effective cost per usage
as a forging hammer, which must be minimized, is predicted using a model of
the weighted commercial elemental prices. The alloys with the most suitable
mechanical properties are expected to be those that possess a Mo solid
solution containing only HfC and other carbide precipitates. The low
diffusion constant in Mo alloys below
1500\degreeC~\cite{Kalinovich65,Imai14} means that the phase stability and
HfC content should reflect the likely room temperature condition of an
as-cast alloy.  The thermodynamic phase stability and HfC content is
evaluated by a neural network trained on a database comprising of {\sc
  calphad} results, with the data sourced from the SSOL6
database~\cite{thermocalcgeneral02,SSOL_2016}. The use of a neural network
to predict phase stability dramatically speeds up the alloy optimization
process as it is computationally less intensive than individual
thermodynamic calculations. It is essential for forging-die alloys to be
strong, particularly in compression, so both the yield stress and also the
hardness must be maximized. However, the yield stress and hardness cannot be
reliably calculated by computer modeling from first principles.  Instead a
database of experimental results for all of the properties as a function of
composition is compiled from the sources referenced in
\tabref{tbl:DataSources} comprising of alloys in an as-cast condition and
exclusively of the Mo solid solution phase behavior prescribed by the
thermodynamic predictions. The scarcity of hardness data means that the
neural network can be improved if it is supplemented with ultimate tensile
strength data. The neural network formalism~\cite{Conduit17} can
automatically identify the link between ultimate tensile strength and
hardness (known to be approximately $\times3$~\cite{Zhang11}) from common
compositions, and then use the surplus ultimate tensile strength data at
other compositions to guide the extrapolation of the hardness model.

After the database of material properties in \tabref{tbl:DataSources} is
compiled, a neural network model is trained on that data to predict the
physical properties for a given composition. The form of neural network and
approach to training follows that in Ref.~\cite{Conduit17} used to develop
Ni-base superalloys.  The design variables were the elemental concentration
of $\{\rm{Nb},\rm{Ti},\rm{C},\rm{Zr},\rm{Hf},\rm{W},\rm{Mo}\}$.  Typically
three hidden nodes gives the best fitting neural network. The neural
  network model predicted not only the expected value of the physical
  property but also the uncertainty associated with it, accounting for
  experimental uncertainty in the underlying data, the uncertainty in the
  extrapolation of the training data~\cite{Heskes97,Papadopoulos01},
and the uncertainty in the processing conditions of as-cast alloys.

In this approach, the individual material properties are converted into a
single merit index that describes the likelihood that the material
properties ($\vec{V}$) satisfies the design criteria ($\vec{T}$) is
$L=\Phi[\vecgrk{\Sigma}^{-1}(\vec{V}-\vec{T})]$. Here $\Phi$ is the
multivariate cumulative normal distribution function and $\vecgrk{\Sigma}$
is the covariance matrix~\cite{Wasserman04}. Combining the individual
property likelihoods enables an estimate to be made of the likelihood that
the alloy will fulfill the whole specification. Critically, this overall
likelihood will be much lower than that of an individual property target
being met. For example, for a five-part specification, if the material has a
$50\%$ likelihood of fulfilling each design criterion, the overall
likelihood that it simultaneously fulfills five criteria is
$0.5^{5}\approx0.03$, so $3\%$. It is therefore crucial that the likelihood
of the material meeting the conformance specification is maximized. The use
of likelihood also allows the tool to explore and select the ideal
compromise between material properties, which is inaccessible with methods
that do not account for likelihood, such as a principal component
analysis~\cite{Toda13} and robust design~\cite{Backman06}.  Similarly, the
design tool may interpolate between experimental data, exploring more
compositions than would be accessible by an experimentally driven
search. Using a neural network to interpolate allows us to capture deeper
correlations than linear regression methods such as those used in principal
component analysis~\cite{Toda13}.

As well as predicting material properties, the tool must simultaneously
optimize them against the set targets.  Previous optimization techniques
included running over a pre-determined grid of compositions, and then
sieving them with trade-off diagrams~\cite{Reed09}, or a Pareto
set~\cite{Bligaard03,Greeley06,Lejaeghere13}. However the expense of these
methods scales exponentially with the number of design variables rendering
them impractical. Another approach is to use genetic
algorithms~\cite{Johannesson02,Stucke03}, but this approach is not
mathematically guaranteed to find the optimal
solution~\cite{Ingber92,Mahfoud95}, and it displays poor performance in high
dimensional problems~\cite{Ingber92,Mahfoud95}. Here we maximize the
logarithm of the likelihood $\log(L)$ to ensure that in the region where the
material is predicted to not satisfy the specification the optimizer runs up
a constant gradient slope that persistently favors the least optimized
property. We explore the high-dimensional composition space with a random
walk which uses a step length that is comparable to the accuracy with which
a material could be manufactured, this is 0.1wt\% for the entire composition
excluding the possibility of microsegregation.  The tool typically search
over $\sim10^8$ sets of design variables in $\sim1$ hour to explore the
space and search for an optimal material.

With the neural network tool in place it is now used to design a new Mo-base
forging die alloy. Once, designed, the properties of the alloy are
subsequently verified by experiment. The goal is to design a new Mo-base
alloy that offers both improved high-temperature hardness and concomitant
greater lifetime with lower in-service costs at $\sim1000${\degreeC}. This
case study not only serves as an independent test of the alloy design
approach, but moreover leads to an alloy with properties that exceed those
of other, commercially available Mo-base alloys.

\begin{figure}[t]
 \centering
 \includegraphics[width=0.8\linewidth]{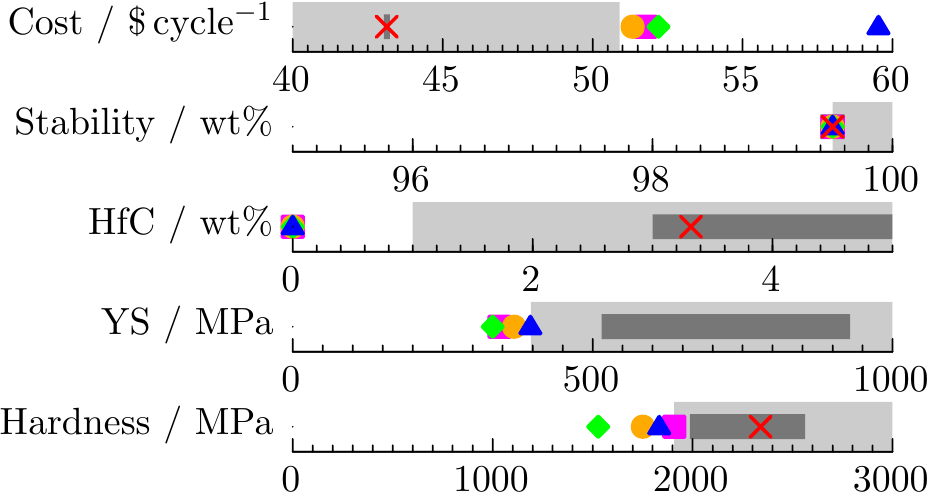}\\[10pt]
 \includegraphics[width=1.0\linewidth]{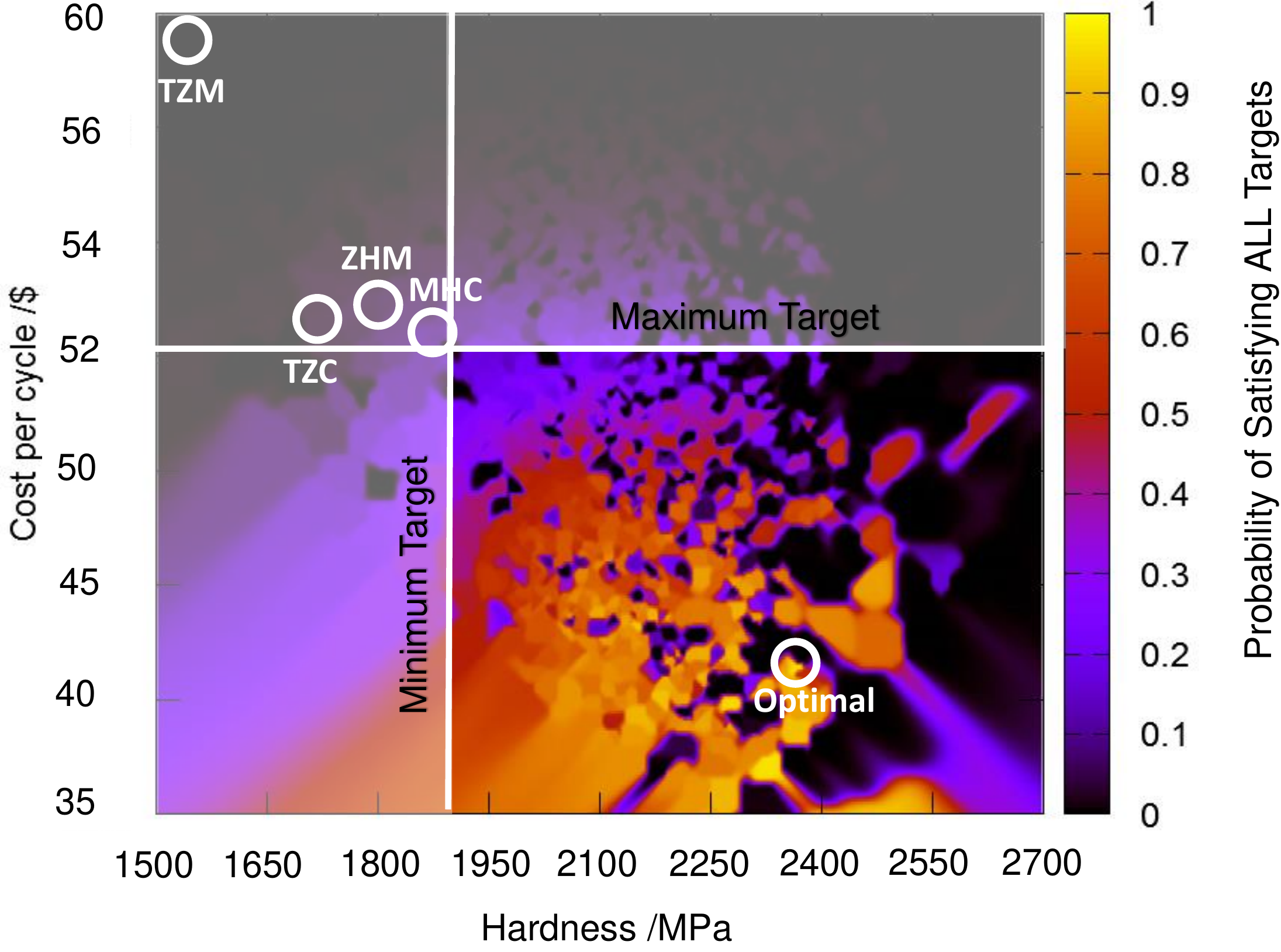}\\
 \caption{{\emph{Upper}:} Summary of properties for the Mo-base alloy. For each listed
   property the gray box refers to the acceptable target properties,
   the dark gray is the three-sigma uncertainty on the theoretical
   prediction.  The points refer to experimentally measured values
   with {\color{red}$\times$}~the proposed alloy,
   {\color{magenta}$\blacksquare$}~MHC, {\color{orange}$\bullet$}~TZC,
   {\color{green}$\blacklozenge$}~TZM, and
   {\color{blue}$\blacktriangle$}~ZHM.
   {\emph{Lower}:} The compromise between hardness and cost per cycle
   made in the design of the Mo-base alloy. The white shaded areas show
   regions that fail to meet hardness and cost targets. The
   color of shading shows the likelihood of exceeding all of the targets,
   following the scale on the right. The white circles show the proposed and
   existing alloys.}
 \label{fig:ResultsMo}
\end{figure}

The first step to design an alloy is to set the target specification. This
is shown in \tabref{tbl:DataSources} and compared with commercially
available Mo-base alloys in \figref{fig:ResultsMo}(a). The alloy should be
cheaper than the previous cheapest Mo forging alloy, TZC, at
$52\,\$\,\rm{cycle}^{-1}$. To avoid forming deleterious phases that could
weaken the alloy it must have good phase stability, defined as the
concentration of the Mo-base solid solution rather than other deleterious
phases, comparable or better than previous Mo alloys of 81wt\%. At the same
time, the Mo alloy should be strengthened by HfC and other carbides, so
there should be at least 1wt\% HfC precipitate formation. The yield stress
should be greater than $398$\,MPa at $1000${\degreeC}, that of the best
alloy available, ZHM. The alloy should also have a hardness higher than the
highest of the Mo alloys, MHC, of $1908$\,MPa at $1000${\degreeC}.  These
targets mean that the new Mo-base alloy will have properties superior to
those of any commercially available alloy. Neural network models for the
cost, phase stability, volume fraction of the reinforcing phase, yield
stress, and hardness are trained using data from the references in
\tabref{tbl:DataSources}. The neural networks will then be used to optimize
the composition to search for the alloy most likely to exceed the target
specification.

\begin{table}
 \small\centering
 \begin{tabular}{ll|ll}
  \multicolumn{4}{l}{Optimal\,composition\,(wt\%)}\\
  \hline
  Nb&\pme{5.7}{0.2}  &Zr&\pme{0.9}{0.1}\\
  Ti&\pme{1.0}{0.1}  &Hf&\pme{9.0}{0.1}\\
  C &\pme{0.20}{0.01}&W &\pme{0.5}{0.2}\\
  Mo&Balance&&
 \end{tabular}
 \caption{The composition of the Mo-base alloy (wt\%). The
   design tolerance shows all compositions that are predicted to fulfill
   the target specification.}
 \label{tab:Composition}
\end{table}

The composition proposed in \tabref{tab:Composition} has a $99.1\%$
likelihood of meeting the target specification in \figref{fig:ResultsMo}(a),
it notably has high levels of Hf at 9wt\% to allow 4wt\% of HfC precipitates
to form, alongside other carbides, to strengthen the alloy. The theoretical
predictions for the alloy all fall within the required targets, with the
alloy being substantially cheaper than required at
$43\,\$\,\rm{cycle}^{-1}$, and the yield stress of $722\,\rm{MPa}$ and
hardness of $2274\,\rm{MPa}$ being considerably greater than the target
specification.  In fact, it is predicted that all of the properties of the
alloy will simultaneously exceed all properties of commercially available
Mo-base alloys.  The composition is quoted with a range of concentrations
that all satisfy the target criteria.

Inevitably, the designer must make a compromise between the different
properties of an alloy. This can be directly visualized,
\figref{fig:ResultsMo}(b) illustrates the trade-off made between hardness
and cost. The positive trend of the bands of iso-likelihood shows how alloys
with good mechanical properties are also more expensive, due to increased
but expensive Hf additives. This landscape allows an engineer to select the
ideal compromise for their application, for example with the aid of an Ashby
plot~\cite{Ashby99}. The proposed alloy is the one most likely to fulfill
the targets highlighted as it lies in the region of highest likelihood. This
is located at a hardness far higher than the minimum hardness target since
there is a large uncertainty on hardness predictions, but nearer to the
maximum cost target since there is a smaller uncertainty on predictions of
cost. The rapidly varying likelihood of satisfying all of the targets
reflects how other properties vary rapidly due to the underlying and locally
optimized composition changing markedly. This variation is similar to that
seen in the design of the Ni-base superalloys~\cite{Conduit17}.

The proposed alloy is predicted to fulfill the target specification.
However, experiments will provide the true test of the performance of the
new alloy.  The synthesis of the proposed Mo-base alloy starts with
pelletized elements having purity greater than $99.9\%$, that are arc-melted
into a $50\,$g ingot through five successive inversion and re-melt cycles.
Brinell hardness testing was conducted on multiple specimens following a 15
minute dwell at the testing temperature. Measurement of the indentation was
obtained using SEM.

\begin{figure}
 \centering
 \includegraphics[width=0.7\linewidth]{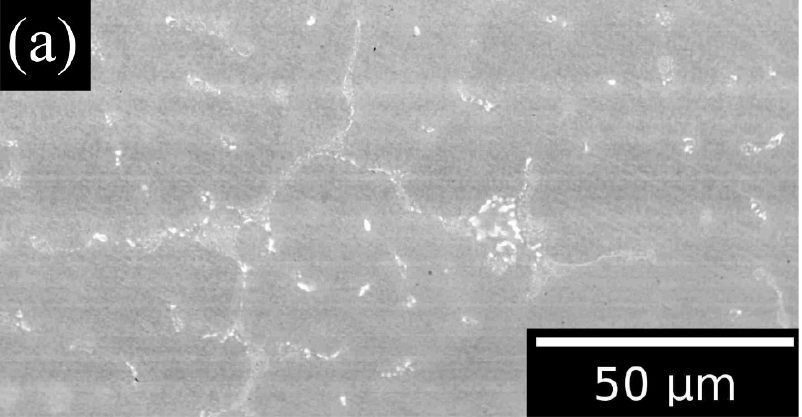}\\[10pt]
 \includegraphics[width=0.7\linewidth]{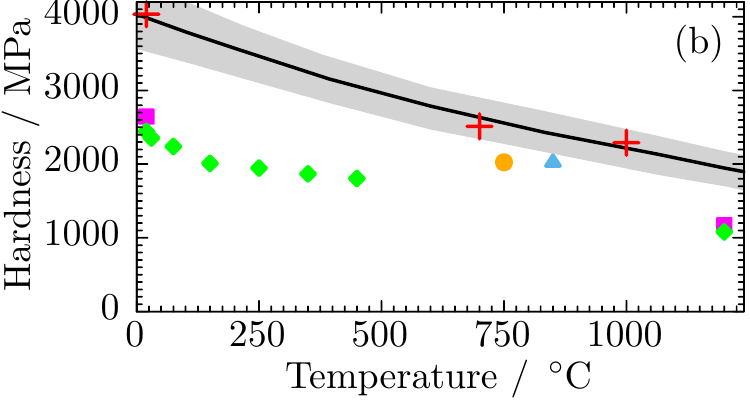}
 \caption{(Color online)
   (a) Secondary electron micrograph image for the Mo alloy.
   (b) Hardness as a function of temperature, the
   black line shows the theoretical prediction and gray the
   uncertainty. The points refer to experimentally measured values
   with {\color{red}$\boldsymbol{+}$}~the optimal alloy,
   {\color{magenta}$\blacksquare$}~MHC, {\color{orange}$\bullet$}~TZC,
   {\color{green}$\blacklozenge$}~TZM, and
   {\color{blue}$\blacktriangle$}~ZHM.
   }
 \label{fig:ExperimentsMo}
\end{figure}

\figref{fig:ExperimentsMo}(a) shows a secondary electron micrograph of the
alloy. The emergence of a single Mo-rich matrix phase strengthened by
carbide precipitates verifies the stability prediction. Spot energy
dispersive X-ray spectroscopy confirmed that the white precipitates are
predominantly HfC, which acts as the main strengthener, with additional
strengthening from Ti, Nb, Ta, and W that are fully miscible above
$882${\degreeC}~\cite{Rudy69} in the Mo-rich solid solution and so minimal
deleterious phases were formed. The fraction of HfC is 4wt\%, in line with
theoretical predictions and greater than that in MHC of 0.5wt\%, so the
alloy should itself have good compressive strength. A trial heat treatment
of 1000{\degreeC} for 20~hours showed no microstructure evolution,
confirming the stability with respect to microstructural evolution. Finally,
the hardness is measured as a function of
temperature. \figref{fig:ExperimentsMo}(b) shows that the alloy possesses a
significantly higher hardness than the commercially available alloys at high
temperatures, making it particularly suitable for refractory applications.

The experimental results are summarized in \figref{fig:ResultsMo}(a). The
four crucial properties of the proposed alloy (cost, phase stability, volume
fraction of the HfC reinforcing phase, and hardness) are in accordance with
the theoretical predictions, exceed the targets, and surpass the properties
of the commercial alloys MHC, TZC, TZM, and ZHM. Furthermore, the neural
network tool has been used to propose another Mo-base
alloy~\cite{Conduit2013iii} but with NbC based hardeners, which has also
been experimentally verified. This both demonstrates the capabilities of the
materials optimization approach and has identified an alloy that may have
potential refractory applications, and in particular as a forging die.

A new computational tool was used to propose the Mo-base alloy most likely
to simultaneously fulfill five different physical criteria given the
experimental and computational data available. The new proposed alloy has
been experimentally verified to have properties that exceed other,
commercially available Mo-base alloys. The Mo-base alloy has the ideal
properties to be used as a forging die for use on future high strength
superalloys at the high temperatures $\sim1000-1100${\degreeC}.

The neural network tool has also been used to design another Mo-based alloy
based on niobium precipitates~\cite{Conduit2013iii}, and two nickel-base
alloys~\cite{Conduit2013ii,Conduit2013iv} that have also been experimentally
verified~\cite{Conduit17}. The capability to rapidly discover materials
computationally should empower engineers to instantly optimize bespoke
materials for their application, bringing materials into the heart of the
design process.

The authors 
acknowledge the
financial support of Rolls-Royce plc, EPSRC under EP/H022309/1 and
EP/H500375/1, the Royal Society, and Gonville \& Caius College. There is
Open Access at
\texttt{https://www.openaccess.cam.ac.uk}.

\bibliography{NeuralData,StdPapers}

\end{document}